\documentclass[prl,showpacs,twocolumn]{revtex4-1}
\usepackage[T1]{fontenc}
\usepackage[latin9]{inputenc}
\usepackage{color}
\definecolor{note_fontcolor}{rgb}{0.80078125, 0.80078125, 0.80078125}
\usepackage{units}
\usepackage{amsmath}
\usepackage{amssymb}
\usepackage{wasysym}
\usepackage{graphicx}

\definecolor{purple}{rgb}{0.85,0.1,0.6}


\begin{document}

\title{Effects of inertia on the steady-shear rheology of disordered solids}

\author{Alexandre Nicolas$^{\S}$}
\altaffiliation[Present address: ]{CAB, 8400 S.C Bariloche, Argentina.}

\author{Jean-Louis Barrat}

\affiliation{LIPhy, Universit\'e Grenoble-Alpes \& CNRS, F-38000 Grenoble,
France}

\author{J\"org Rottler$^{\S}$ ($^{\S}$: equal contribution)}

\affiliation{Department of Physics and Astronomy, The University of British Columbia,
6224 Agricultural Road, Vancouver, British Columbia V6T 1Z4, Canada}

\begin{abstract}
We study the finite-shear-rate rheology of disordered solids by means
of molecular dynamics simulations in two dimensions. By systematically
varying the damping strength $\zeta$ in the low-temperature limit, we
identify two well defined flow regimes, separated by a thin
(temperature-dependent) crossover region.  In the overdamped regime,
the athermal rheology is governed by the competition between elastic
forces and viscous forces, whose ratio gives the Weissenberg number
$\mathrm{Wi}\propto\zeta\dot{\gamma}$; the macroscopic stress $\Sigma$
follows the frequently encountered Herschel-Bulkley law
$\Sigma=\Sigma_{0}+k\sqrt{\mathrm{Wi}}$, with yield stress
$\Sigma_{0}>0$. In the underdamped (inertial) regime, dramatic changes
in the rheology are observed for low damping: the flow curve becomes
nonmonotonic. This change is not caused by longer-lived correlations
in the particle dynamics at lower damping; instead, for weak
dissipation, the sample heats up considerably due to, and in
proportion to, the driving. By thermostatting more or less underdamped
systems, we are able to link quantitatively the rheology to the
kinetic temperature and the shear rate, rescaled with Einstein's
vibration frequency.
\end{abstract}


\maketitle 

Inertia matters in liquid flows. Its presence in the Navier-Stokes
equations leads to a rich phenomenology that vanishes in the
overdamped limit of viscous flow. However, the effect
of damping is rarely heeded (let alone analyzed) in the flow of
disordered solids, so much so that dense colloidal glasses often serve
as model systems for bulk metallic glasses (BMG) \cite{amann2013stress}, even
though they are much more strongly damped. Here, we find
that reducing the damping  can dramatically impact the
macroscopic rheology.  In the inertial regime,  the energy input dwells
longer in the particle momenta before its final dissipation into the heat bath,
thus
facilitating plastic flow. We provide a quantitative account of this
effect in terms of simple kinetic heating of the underdamped solid,
similar to the one observed experimentally during the operation of
shear bands in BMG
\cite{Lewandowski2006temperature,Zhang2007local}.

The damping regime is not the only line of contrast among disordered
solids: atoms in BMG as well as
small colloids are heavily influenced by thermal fluctuations whereas
grains are quasi-athermal; foam bubbles are
deformable whereas some colloids are close to perfect hard
spheres. Notwithstanding these contrasting features, virtually all
such solids deform similarly, \emph{i.e.}, mostly elastically at
small stresses while at larger shear plasticity becomes dominant, with
a succession of failures of micro-regions, whose particles rearrange
swiftly. These rearrangements are triggered by the loading or
facilitated by thermal activation
\cite{Schall2007,Maloney2006Amorphous} and may interact \emph{via} the
long-range elastic deformation that they induce in the surrounding
medium \cite{Chattoraj2011robustness,Maloney2006Amorphous}.  Based on
this generic scenario, multiple simplified rheological models have
been proposed, generally focusing on the overdamped regime
\cite{Bulatov1994,Falk1998,Sollich1997,Hebraud1998,Picard2005,Fuchs2002}
(nevertheless, the mesoscale elastic response has been studied across
the damping regimes
\cite{Puosi2014,Nicolas2015elastic}).
To what extent does the presence of inertia alter the picture?

In the quasistatic limit, \emph{i.e.}, at vanishing shear rates
$\dot\gamma$, recent numerical work by Salerno and Robbins has
ascertained that the statistics of avalanches fall into\emph{
 distinct} universality classes in the overdamped \emph{vs.} underdamped regimes
\cite{Salerno2012,Salerno2013effect}.  The
difference is best illustrated by considering the complex, rugged
Potential Energy Landscape (PEL) in which the system evolves: it
climbs up energy barriers in the phases of elastic loading and
abruptly slides downhills once the barrier is overcome. For overdamped
systems, this descent suffices to dissipate the energy stored during
loading, while at lower damping the inertial force 
may carry the system over several successive barriers. This process is then
highly directional in the PEL and strongly correlated in
space and time, which renders its modelling quite complex \emph{a
  priori}. Some \emph{ad hoc} rules to include inertia in
lattice-based models have been put forward, such as lowering barriers
or yield stresses for a certain time after failure (see
\cite{Prado1992inertia} and references in \cite{Salerno2013effect})
and their impact has been emphasized, but the validity of these
descriptions stands on shaky ground.

In this Letter, we focus on the steady-state shear flow of
two-dimensional disordered solids at \emph{finite} driving rates and
investigate the role of inertia in the vanishing and low temperature
limits, with molecular dynamics (MD) simulations. We apply simple
shear to the binary Lennard-Jones glass used in
Ref.~\cite{Nicolas2015elastic}; it comprises 32,500 large (type A) particles
and 17,500 small (type B) particles, all of mass $m$, and has reduced
density $\rho=1.2$.  The equations of motion are based on the
Dissipative Particle Dynamics (DPD) scheme \cite{Soddemann2003} and
read
\begin{eqnarray}
\begin{cases}
\frac{d\mathbf{r_{i}}}{dt} & ={\bf v_{i}}\\
m\frac{d\mathbf{v_{i}}}{dt} & =-\sum_{i\neq
j}\frac{\partial\mathcal{V}\left(r_{ij}\right)}{\partial\mathbf{r_{ij}}}
+\boldsymbol{f_{i}}^{R}+\boldsymbol{f_{i}}^{D}.
\end{cases}\label{eq:eq_of_motion}
\end{eqnarray}
Here, $\boldsymbol{r_{ij}}\equiv\boldsymbol{r_{i}-r_{j}}$ and
$\mathcal{V}(r_{ij})$ is the interaction potential between particles $i$ and
$j$. The DPD
forces  $\boldsymbol{f_i}^{R,D}$ involve a cutoff
function $w(r) \equiv 
1-\frac{r}{r_{c}}$ if $r<r_{c}\equiv2.5\sigma_{AA}$ and 0
otherwise; $\boldsymbol{f_{i}}^{R}  \equiv  s\sum_{j \neq 
i}w\left(r_{ij}\right) \theta_{ij}\frac{\boldsymbol{r_{ij}}}{r_{ij}}$ is a
stochastic force, based on the Gaussian white noise
$\theta_{ij}$ \cite{Soddemann2003} and
due to the coupling
to a heat bath maintained at temperature
$T_{0}$ and
$\boldsymbol{f_{i}}^{D}  \equiv -\zeta\sum_{j\neq
i}w^{2}\left(r_{ij}\right)\frac{\boldsymbol{v_{ij}}\cdot\boldsymbol{r_{ij}}}{r_{
ij}^{2}}\boldsymbol{r_{ij}}$
is a damping
force depending on the relative velocities
$\boldsymbol{v_{ij}}\equiv\boldsymbol{v_{i}}-\boldsymbol{v_{j}}$.
The strength $s=2\zeta k_{B}T_{0}$ of the coupling to the reservoir depends on
the damping strength $\zeta$
and $T_{0}$, and is maintained even if the
system departs from thermal equilibrium. In the following, $\zeta$, $m$, and
$T_{0}$ shall be varied, while the particle
interactions are kept constant.

\smallskip
The equations of motion, Eqs.~\ref{eq:eq_of_motion}, are integrated on GPU with
the velocity Verlet algorithm. They involve forces
deriving from four types of stresses:

(i) the elastic stress, of order
$\Sigma_{A}\equiv\frac{\epsilon_{AA}}{\sigma_{AA}^{2}} \equiv 1$,

(ii) the viscous stress, of order $\eta\dot{\gamma},$ where $\eta\approx\zeta$
\cite{Nicolas2015elastic} is the microscopic viscosity,

(iii) the inertial pressure, which, in a Bagnold-like picture
\cite{Bagnold1954experiments},
involves momentum transfers of order $m\sigma_{AA}\dot{\gamma}$ at a rate
$\propto\dot{\gamma}$, and is thus proportional
to $m\dot{\gamma}^{2}$, and 

(iv) the thermal pressure resulting from stochastic forces of magnitude
$\sqrt{\zeta T_{0}}$.

Their relative magnitudes are quantified by dimensionless numbers
that characterize the flow regime. In particular, the importance of
viscosity with respect to elasticity is measured by the Weissenberg
number,
\[
\mathrm{Wi}\equiv\tau_{\mathrm{diss}}\dot{\gamma}\text{ with
}\tau_{\mathrm{diss}}\equiv\frac{\zeta}{\Sigma_{A}},
\]
and the ratio of inertial over elastic stresses is given by $\mathrm{Ei}^{2}$,
where
\[
\mathrm{Ei}\equiv\tau_{\mathrm{vib}}\dot{\gamma}\text{ with
}\tau_{\mathrm{vib}}\equiv\sqrt{\frac{m}{\Sigma_{A}}}.
\]

In conjunction with $T_{0}$, Wi and Ei fully characterize the flow.
Nevertheless, to describe the damping regime of flow curves, irrespective of the
shear rate, it is convenient to also introduce
\[
\mathrm{Q}\equiv\frac{\mathrm{Ei}}{\mathrm{Wi}}=\frac{\sqrt{m\Sigma_{A}}}{\zeta}
= \frac{\tau_{\mathrm{damp}}}{\tau_{\mathrm{vib}}} \text{ with }
\tau_{\mathrm{damp}}\equiv \frac{m}{\zeta};
\]
if Eq.~\ref{eq:eq_of_motion} is assimilated to a damped second-order
harmonic oscillator, Q is the (inertial) quality factor, \emph{i.e.}, the number
of inertial oscillations in the damping time.

\begin{figure}[t]
\begin{centering}
\includegraphics[width=8cm]{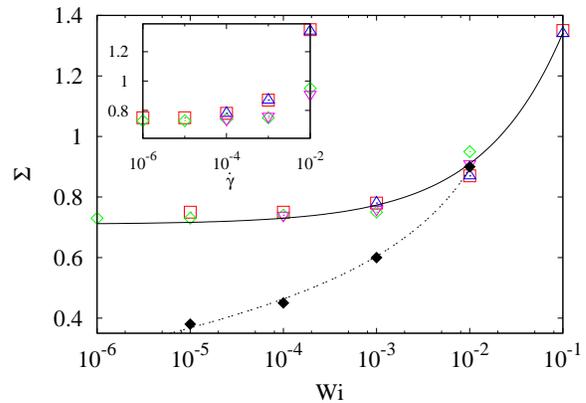}
\caption{\label{fig:Flow-curves-BD} Athermal flow curves
$\Sigma(\mathrm{Wi},T_{0}=0)$ in the overdamped regime
$\mathrm{Q}\lesssim1$, for various combinations $[\zeta,m]$: $[\zeta=1,\ m=1]$
($\lozenge$), $[10,1]$ ($\square$), $[10,0.1]$  ($\triangle$),
and $[1,0.1]$ ($\triangledown$). The solid line represents
Eq.~\ref{eq:HB_overdamped}. A flow curve at $T_{0}=0.2$,
$\mathrm{Q}=1$ $[1,1]$ ($\blacklozenge$) is also shown. The
thin dashed line is a best fit to 
 Eq.~\ref{eq:CCL}. Inset: $\Sigma$ \emph{vs.} $\dot\gamma$.}
\end{centering}
\end{figure}

\begin{figure}
\begin{centering}
\includegraphics[width=8cm]{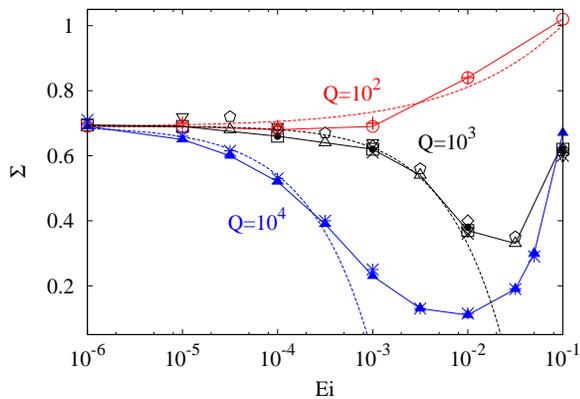}
\par\end{centering}
\caption{\label{fig:thermostatted_flow_curves}
Flow curves $\Sigma(\mathrm{Q},\mathrm{Ei},0)$ of athermal underdamped
systems and
$\Sigma\left(\mathrm{Q^\prime},\mathrm{Ei},T_{K}\left(\mathrm{Q},\mathrm{Ei}
\right)\right)$ of their thermostatted counterparts (\emph{see text}). Symbols
are listed in Table~\ref{table}. Thin dashed lines are the best fit to
Eq.~\ref{eq:CCL},
$\Sigma=0.69+$ $2 \sqrt{\mathrm{Ei}}-0.17T_{K}^{\nicefrac{2}{3}}\ln\left(\frac{
0.4T_ { K }
^{\nicefrac{5}{6}}}{\mathrm{Ei}}\right)^{\nicefrac{2}{3}}$,
where $T_{K}=0.15\mathrm{Q}\cdot\mathrm{Ei}$. }
\end{figure}

Our numerical data confirm the relevance of such dimensional analysis:
Figures~\ref{fig:Flow-curves-BD}
and \ref{fig:thermostatted_flow_curves} prove that, at
$T_{0}=0$, the dependences of the macroscopic shear stress $\Sigma$ on
$\zeta$, $m$, and $\dot{\gamma}$ can be condensed into a dependence
on the pair $\left(\mathrm{Q},\mathrm{Wi}\right)$, or equivalently
(but more conveniently when $\mathrm{Q}\gg1$)
$\left(\mathrm{Q},\mathrm{Ei}\right)$.

\paragraph*{Overdamped dynamics.}

Let us start by investigating the fully overdamped (Brownian or athermal) limit
$\mathrm{Q}\rightarrow0$. In the absence of inertia, $\dot{\gamma}$ is best
rescaled as Wi. At $T_{0}=0$, the flow
curve, plotted in Fig.~\ref{fig:Flow-curves-BD}, is very well described
by the Herschel-Bulkley law
\begin{equation}
\Sigma(\mathrm{Wi},T_{0}=0)=0.72+2\,\sqrt{\mathrm{Wi}}.
\label{eq:HB_overdamped}
\end{equation}
Interestingly, this description remains very good at finite values
of Q, up to $\mathrm{Q}\approx 1$.
Thus, for all $\mathrm{Q}\leqslant1$,
the macroscopic rheology is exclusively governed by the competition
between elastic and viscous forces.

Leaving the athermal regime, we observe that imposing a finite bath
temperature $T_{0}>0$ leads to a decrease of $\Sigma$ at all shear
rates.  Regardless of the damping regime, this thermal effect is
explained by the premature occurrence of plastic rearrangements owing
to thermal activation: the system at $T_{0}>0$ hops over saddle-node
points in the PEL before the effective potential barriers have
completely flattened under the influence of the driving, which interrupts the
elastic accumulation of strain; hence the lower
macroscopic stress \cite{Johnson2005,Chattoraj2010}.

Perhaps less expectedly, we also find a narrowing of the
overdamped regime with $T_0$, that is, the quality factor
$\mathrm{Q}_{c}(T_{0})$ marking the departure from the scaling with Wi
decreases with $T_{0}$ (our data suggest
$\mathrm{Q}_{c}\left(T_{0}=0\right)\approx1$ whereas
$\mathrm{Q}_{c}\left(T_{0}=0.2\right)<1$ but do not allow for greater
accuracy). A rather general explanation consists in alluding to the
excitation of higher-frequency modes at higher temperature, these
modes having larger specific quality factors Q, or to the faster
thermalization of the system (\emph{see below}).

\paragraph*{Inertial dynamics.}

On increasing Q, past a small crossover region around $\mathrm{Q_{c}}(T_{0})$,
one enters the underdamped regime, where the rheology is \emph{a priori}
described by the triplet $\left(\mathrm{Q},\mathrm{Ei},T_{0}\right)$.
What role does the inertial quality factor Q play in that regime? 

In fact, at low damping, Q can no longer be interpreted as
the number of not-too-damped inertial oscillations within a particle's
cage. Indeed, localized
excitations spread in the glass and, owing to
nonlinearities, thermalize: their energy is
redistributed across the whole vibrational spectrum. This process
occurs over a time $\tau_{\mathrm{th}}$ and expedites the
decorrelation of the excitations when the velocity damping time
$\nicefrac{m}{\zeta}$ becomes longer than $\tau_{\mathrm{th}}$. As a
result, the velocity autocorrelation functions, which reflect
single-particle dynamics, gain independence from Q, in the quiescent
system at $T_{0}>0$ (in Fig.~1 of the
Supplementary Material (SM), we observe
$\tau_{\mathrm{th}}\approx0.1\tau_{\mathrm{vib}}$ at $T_{0}=0.16$).
Thus, one is lured into thinking that the underdamped rheology is
insensitive to Q, in the same way as the equilibrium properties of
liquids computed with MD are independent of the (weak) damping
\cite{Evans1984equilibrium,Soddemann2003}.

\begin{table}

\begin{tabular*}{0.45\textwidth}{ @{\extracolsep{\fill} }  c  c  c  c  c  c
 }
  \hline
  $\mathrm{Q}$ & $\mathrm{Q}^\prime$ & $ \zeta $ & $m$ & $T_0$ & symbol  \\
  \hline
   $10^2$ &     & $10^{-2}$ &  1  &      0       &     $\circ$      \\
          & 10  &  0.1      &  1  & $T_K(\mathrm{Q}=10^2,\mathrm{Ei})$ &    
$+$   \\
   $10^3$ &     & $10^{-3}$ &  1  &      0       &     $\square$      \\
   $10^3$ &     & $10^{-2}$ &  100  &      0       &     $\bullet$      \\
   $10^3$ &     & $3\cdot10^{-4}$ &  0.1  &   0   &     $\triangle$   \\
             & 1  &  1      &  1  & $T_K(\mathrm{Q}=10^3,\mathrm{Ei})$ &    
$\triangledown$   \\
             & 1  &  10      &  100  & $T_K(\mathrm{Q}=10^3,\mathrm{Ei})$ &    
$\lozenge$   \\
            & 1  &  0.3      &  0.1  & $T_K(\mathrm{Q}=10^3,\mathrm{Ei})$ &    
$\pentagon$   \\
          & 10  &  0.1      &  1  & $T_K(\mathrm{Q}=10^3,\mathrm{Ei})$ &    
$\times$   \\
  $10^4$ &     & $10^{-4}$ &  1  &   0   &     $\blacktriangle$   \\
            & 1  &  1      &  1  & $T_K(\mathrm{Q}=10^4,\mathrm{Ei})$ &    
$\ast$   \\
\hline
\end{tabular*}

\caption{\label{table}Parameters and symbols used in Figs.
\ref{fig:thermostatted_flow_curves} and
\ref{fig:Kinetic-temperature-measured}. Inertial quality factors are
denoted $\mathrm{Q}^{\prime}$, instead of $\mathrm{Q}$, for the
thermostatted systems (see text).}
\end{table}

Contrary to this thought, the underdamped flow curves,
plotted in Fig.~\ref{fig:thermostatted_flow_curves}, exhibit dramatic
changes at large Q (and low $T_{0}$), as they become nonmonotonic!

\begin{figure}
\begin{centering}
\includegraphics[width=8cm]{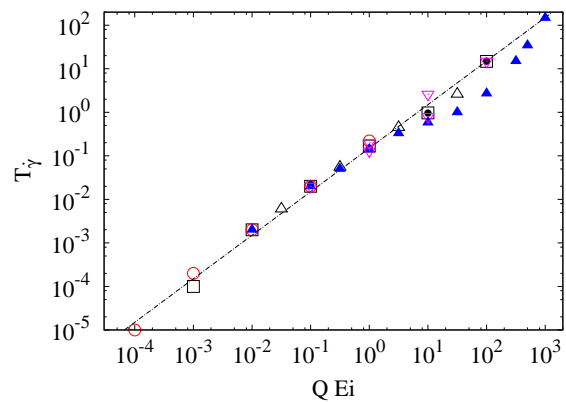}
\par\end{centering}
\caption{\label{fig:Kinetic-temperature-measured} Shear contribution
$T_{\dot\gamma}= T_K - T_0$ to the kinetic temperature
 measured in underdamped samples
  \emph{vs.} $\mathrm{Q}\,\mathrm{Ei}$. \\($\color{purple}\triangledown $) Data
at
$T_0=0.2$. The line represents
$T_{\dot\gamma}=0.15\cdot\mathrm{Q}\cdot\mathrm{Ei}$.}
\end{figure}

Clearly, the insensitivity to Q was a fallacy. In fact, this parameter
also controls energy dissipation in the system. When the damping is
too weak compared to the energy input, the system heats up and strongly
departs from thermal equilibrium with the heat reservoir at $T_{0}$.
This is not a numerical
artifact: in experiments on sheared granular matter, the ``granular
temperature'' differs from room temperature \cite{Losert2000}; temperature
rises have also been borne out experimentally in shear bands in compressed
BMG
\cite{Lewandowski2006temperature,Zhang2007local} (incidentally,
note that a negative rate-dependence of the stress,
known as ``rate-weakening'', has also been reported in these materials
\cite{Dubach2007deformation}). For ``dry'' systems, heat is actually
removed \emph{faster} in simulations than in experiments, where its
extraction must proceed through the boundaries \cite{Bailey2006atomistic}.
Besides, nonmonotonic flow curves are not a marginal effect of
the DPD thermostat; they were also observed by Salerno with a
weak Langevin thermostat (see Fig.~2.3 of Ref.~\cite{Salerno2013thesis}).

Taking into account the heating of the sample, we propose to substitute,
in the triplet $(\mathrm{Q},\mathrm{Ei},T_{0})$, the reservoir temperature
$T_{0}$ with the actual kinetic temperature of the sample,
$T_{K}\equiv\frac{1}{2N}\sum_{i=1}^{N}mv_{i}^{2}$. To assess the
contribution $T_{\dot{\gamma}}$ of the driving to $T_{K}$, we assume
that the kinetic energy is mainly generated by plastic rearrangements, during
which the elastic energy $\frac{1}{2\rho}\Sigma_{0}\gamma_{y}$ per particle
is first converted into kinetic energy and then gradually dissipated,
over a timescale $\tau_{\mathrm{damp}}\equiv\nicefrac{m}{\zeta}$ in the
underdamped
regime. Thus, the density of simultaneous events is
$\nicefrac{m}{\zeta}\cdot\nicefrac{\dot{\gamma}}{\gamma_{y}}$,
and we arrive at
\begin{eqnarray*}
T_{\dot{\gamma}} & \approx &
\left(\frac{1}{2\rho}\Sigma_{0}\gamma_{y}\right)\frac{m\dot{\gamma}}{
\zeta\gamma_ {y}
}\approx\frac{\Sigma_{0}}{2\rho}\cdot\mathrm{Q}\cdot\mathrm{Ei}.
\end{eqnarray*}
The scaling law with $\mathrm{Q}\cdot\mathrm{Ei}$ is in very good agreement with
the numerical data
at $T_{0}=0$, as shown in Fig.~\ref{fig:Kinetic-temperature-measured},
as long as $\mathrm{Q}\gg1$ (see Section V of SM for a
tentative rationalization of the slight deviations). 
The predicted (0.29) and measured (0.15) prefactors differ by a factor of 2
precisely, probably because the released energy is actually equipartitioned
between the kinetic and elastic degrees of freedom, as for a harmonic
oscillator.
Furthermore, we observe a Boltzmann distribution, parametrized
by $T_{\dot{\gamma}}$, of kinetic
energies among the particles (Fig.~2 of SM),
which confirms the
status of $T_{\dot{\gamma}}$ as the sample temperature. This is consistent with
the ``quasi-equilibrium'' situation (at $T_K$) reported by Xu \emph{et
al.} in strongly-sheared athermal systems \cite{Xu2005effective}.
At finite $T_{0}$, we expect
$T_{K}\approx T_{0}+T_{\dot{\gamma}}$,
which is entirely compatible with our (limited) dataset
(Fig.~\ref{fig:Kinetic-temperature-measured}).

Coming back to the underdamped flow curves,
Fig.~\ref{fig:thermostatted_flow_curves}
gives ample evidence that the athermal flow curves
at any $\mathrm{Q}\gg\mathrm{Q}_{c}(0)$ can be quantitatively
reproduced by thermostatting a less underdamped (but still inertial)
system, at $\mathrm{Q}^{\prime}<\mathrm{Q}$, to the shear-rate-dependent
temperature $T_{K}\left(\mathrm{Q},\mathrm{Ei}\right)$ of the original
system; this holds true at $T_{0}>0$ (\emph{data at $T_{0}=0.2$
not shown}). Put differently,
$\Sigma(\mathrm{Q},\mathrm{Ei},T_{0})$ collapses onto a master curve
$\tilde\Sigma\left(\mathrm{Ei},T_{K}\left(\mathrm{Q},\mathrm{Ei}\right)\right)$,
irrespective of the value of Q. Thus, Q does not impact the
underdamped rheology as the inertial quality factor,
but only \emph{via} its control of
$T_{K}\left(\mathrm{Q},\mathrm{Ei}\right)$.
It follows that inertial vibrations and thermal fluctuations have
an analogous effect on the rheology: both are ``agitation'' forces
that precipitate rearrangements, but the former increase with the shear rate,
hence the severe rate-weakening observed in strongly underdamped systems.
We should mention that rate-weakening is generally associated with a flow
instability leading to shear-banding
\cite{Yerushalmi1970stability,Spenley1996nonmonotonic}, but here we have not
seen any banding of the velocity profiles. We believe that this is due to the
rapidity of equilibration through thermal diffusion in small systems, which
impedes the coexistence of bands sheared at different rates,
thus (here) at distinct temperatures.

Chattoraj {\it et al.} \cite{Chattoraj2010}, building on previous
work by Johnson and Samwer \cite{Johnson2005}, propounded the following formula
for the temperature dependence of the stress,
\begin{eqnarray}
\Sigma(\dot{\gamma},T_{0}) & = &
\Sigma(\dot{\gamma},T_{0}=0)-AT^{\nicefrac{2}{3}}\ln\left(\frac{BT_{0}^{
\nicefrac{5}{6}}}{\dot{\gamma}}\right)^{\nicefrac{2}{3}},
\label{eq:CCL}
\end{eqnarray}
where $A$ and $B$ are adjustable parameters. Substituting $T_{0}$
with $T_{K}= T_{0}+0.15\mathrm{Q}\cdot\mathrm{Ei}$
and $\dot{\gamma}$ with Ei in Eq.~\ref{eq:CCL}, we obtain
predictions in broad agreement with our data, as shown in
Fig.~\ref{fig:thermostatted_flow_curves},
as long as the flow remains underdamped and at low enough $T_{K}$.

These results do not imply that in underdamped systems inertia
can be discarded in favor of temperature. Indeed, the collapse onto
$\tilde\Sigma\left(\mathrm{Ei},T_{K}\left(\mathrm{Q},\mathrm{Ei}\right)\right)$
breaks down for $\mathrm{Q}<\mathrm{Q}_{c}(T_{0})$, which highlights
the operativeness of an inertial mechanism at
$\mathrm{Q}>\mathrm{Q}_{c}(T_{0})$,
responsible \emph{e.g.} for the scaling of the inverse attempt frequency
(multiplied by $\dot\gamma$) with Ei, and not Wi. Still, it
is noteworthy that the collapse holds down to values of Q
in the crossover region; in particular, systems at $\mathrm{Q}=1$
display a macroscopic rheology close to the fully overdamped one at
$T_{0}=0$, while a scan through their higher-temperature response
gives access to the strongly underdamped rheology.

In summary, the variations of the macroscopic
rheology of a model
disordered solid with damping strength $\zeta$ (or particle mass
$m$) can be collapsed into two flow regimes.  When
$\mathrm{Q}\equiv\mathrm{Ei}/\mathrm{Wi}$ is smaller than a
threshold $\mathrm{Q}_{c}(T_{0})$, the system is overdamped. It is
widely accepted that foams, concentrated emulsions, and dense
colloidal suspensions belong in this regime. At fixed $T_{0}$, in
particular $T_{0}=0$, the flow curves only depend on Wi, which proves
that the competition between the elastic interactions imposed by the
PEL and dissipation forces dominates the rheology of these
systems. This is compatible with the rheological models proposed by us
and others in
\cite{Hebraud1998,Picard2005,Nicolas2014rheology,Agoritsas2015}, but rules
out all explanations based on the
transverse sound velocity $c_{s}$ (which affects Ei and Q,
but not Wi).

Such explanatory scenarios based on $c_{s}$ could be valid
in the (moderately) underdamped regime, at
$\mathrm{Q}\gtrsim\mathrm{Q}_{c}(T_{0})$.
As a noteworthy example, Lemaître and Caroli \cite{Lemaitre2009rate}
suggested the following scenario, later taken up and revised in
\cite{Fusco2014rheological,Lin2014scaling}
: avalanches of plastic events spread at speed
$c_{s}$ and their spreading is limited by the driving, which generates
independent plastic events. The ensuing incomplete plastic relaxation
explains the increasing flow curve. However, the athermal MD
system used in \cite{Lemaitre2009rate} appears similar to ours
with $\mathrm{Q}\approx 0.2<\mathrm{Q}_{c}(T_{0}=0)$.

For even more strongly underdamped systems, at $\mathrm{Q}\gg10^{2}$, the flow
curve becomes nonmonotonic at
low bath temperature. Surprisingly, this transition has never
been analyzed before, although the threshold for $\mathrm{Q}$
does not seem unrealistically large: a crude estimate for
a suspension of frictionless grains (of density $\rho$ and radius
$a$) in a solvent of viscosity $\eta$ gives
$\mathrm{Q}\approx \frac{0.1a\sqrt{\rho\Sigma_{0}}}{\eta}$. We
showed that variations in the inertial properties of the material
played no role \emph{per se} in the transition; instead, the latter originates
from the insufficient energy dissipation at large Q, which causes
the sample to heat up (and hence, relax stress) all the more as the
driving is fast, with the scaling
law $T_{\dot{\gamma}}\propto\mathrm{Q}\cdot\mathrm{Ei}$. 

This rate-weakening mechanism persists until it is counterbalanced by the
standard collisional increase of $\Sigma$ at high rates; this results in a
minimum in the flow curve (see Section IV of SM for a discussion). The
mechanism is reminiscent of the one producing a shear-banding
instability in the Soft Glassy Rheology (SGR)'s variant proposed by
Fielding \emph{et al}. \cite{Fielding2009}. In SGR, material subunits
possess a (widely distributed) energy barrier for yielding, which decreases as
the material is loaded. Yielding is then
activated by an effective mechanical temperature
$x$. In the variant of Ref.~\cite{Fielding2009}, $x$ is coupled
to the local plastic activity and thus increases with the shear rate.
In a similar fashion, in the Shear Transformation Zone theory, the
strain may localize \emph{via} a coupling between the strain rate
and the ``configurational disorder temperature'' \cite{Manning2007}.
The major conceptual divergence between these approaches and our observations
in severely underdamped systems is the (effective or kinetic) nature
of the temperature.

This difference echoes a vast debate in the metallic glass community
regarding the origin of the softening of shear bands: does the band
persist by softening because of heat production, hence, higher local
temperatures, or, perhaps more probably, because of local configurational
changes (in free volume or density), while the temperature rise is
but a side-effect
\cite{Lewandowski2006temperature,Zhang2007local,Bailey2006atomistic}?
Our findings do not contribute to settling this question, but they do certainly
call for a clarification of the description of damping in
rheological models.

\begin{acknowledgments}
\paragraph*{Acknowledgments.}
We are grateful to Mark Robbins for first mentioning the nonmonotonic
inertial flow curves, we acknowledge discussions with Kirsten Martens, Kamran
Karimi, and Claus Heussinger. The simulations were carried out using the LAMMPS
molecular dynamics software. JLB is supported by Institut Universitaire
de France and by grant ERC-2011-ADG20110209. 

\end{acknowledgments}

\end{document}

\title{Supplemental Material: Effects of inertia on the steady-shear rheology of
disordered solids}

\author{Alexandre Nicolas$^{\S}$, Jean-Louis Barrat}

\affiliation{LIPhy, Universit\'e Grenoble-Alpes \& CNRS, F-38000 Grenoble,
France}

\author{J\"org Rottler$^{\S}$ ($^{\S}$: equal contribution)}

\affiliation{Department of Physics and Astronomy, The University of British
Columbia,
6224 Agricultural Road, Vancouver, British Columbia V6T 1Z4, Canada}

\maketitle

\section{I. Velocity autocorrelation functions}

\begin{figure}[H]
\begin{centering}
%\subfloat[$\dot{\gamma}=0$ and $T_{0}=0.16$.]
{\begin{centering}
\includegraphics[width=8cm]{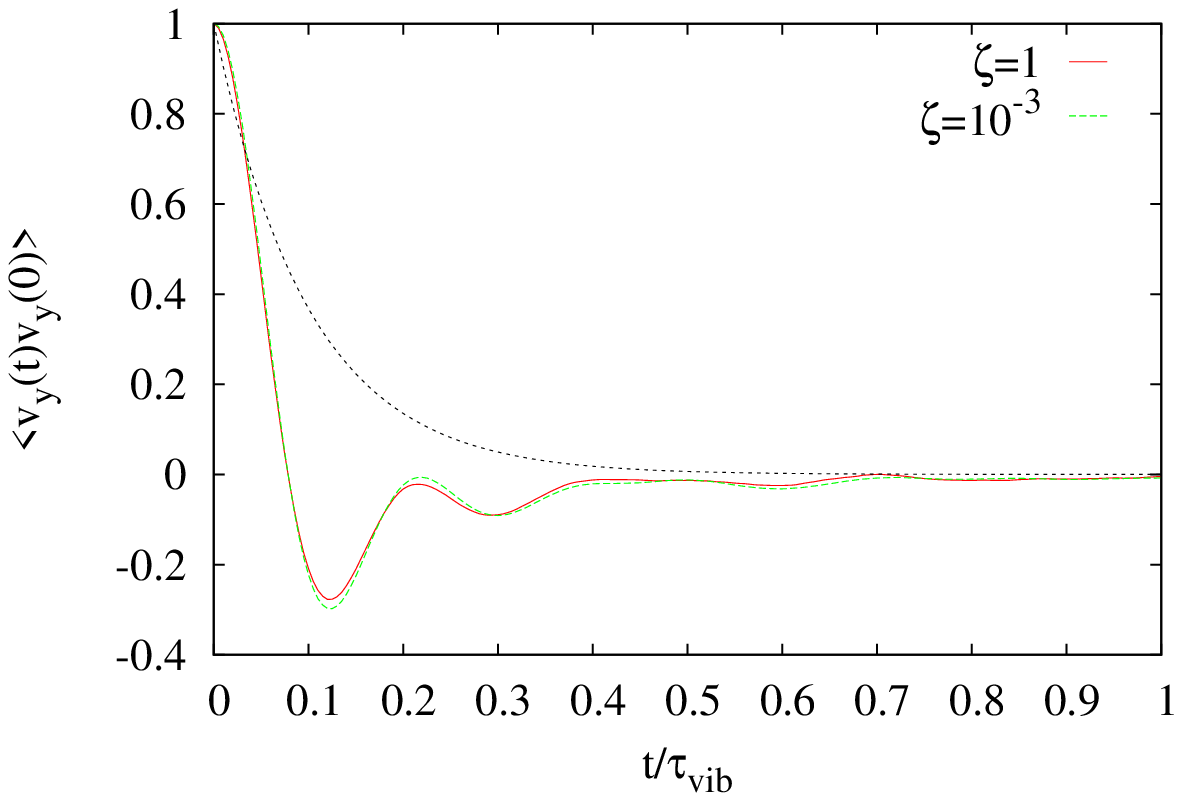}
\par\end{centering}

}
\par\end{centering}

\begin{centering}
{\begin{centering}
\includegraphics[width=8cm]{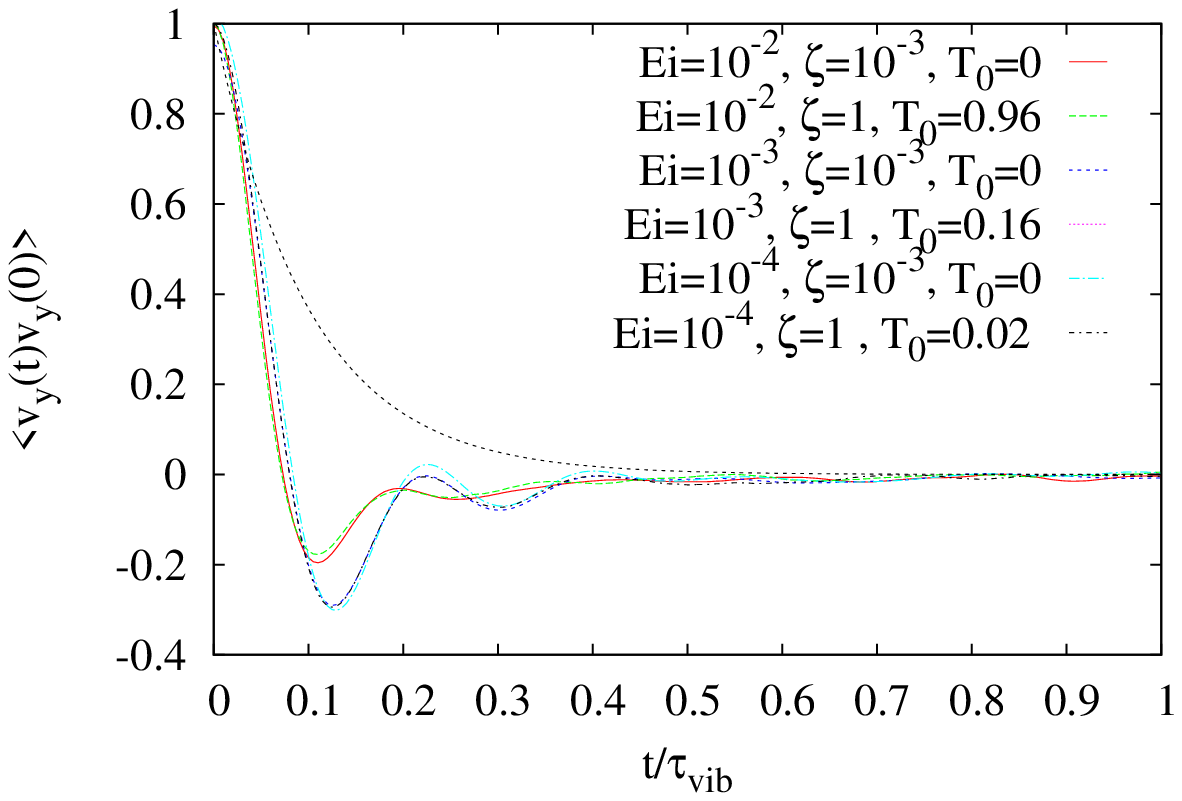}
\par\end{centering}

}
\par\end{centering}
\begin{centering}

\caption{ \label{fig:vacf_quiescent}Transverse velocity
autocorrelation functions for particle mass $m=1$ and different damping
magnitudes $\zeta$ in (top) quiescent systems, \emph{i.e.}, $\mathrm{Ei}=0$, at
$T_{0}=0.16$ and
(bottom) sheared systems, with  Ei and $T_{0}$ as
indicated in the legend. Note that the values of $T_{0}$ used for the flows at
$\zeta=1$ were chosen to match the kinetic temperatures of the strongly
underdamped flows ($\zeta=10^{-3}$) at the same rescaled shear rate Ei. The
dashed black line is an envelope fit with the function
$\mathrm{exp}(\frac{-t}{0.1\tau_\mathrm{vib}})$}
\end{centering}
\end{figure}

Figure~\ref{fig:vacf_quiescent} presents the transverse velocity
autocorrelation functions for different damping strengths, in the quiescent
system.

\section{II. Kinetic energy distributions}

\begin{figure}[H]
\begin{centering}
\includegraphics[width=9cm]{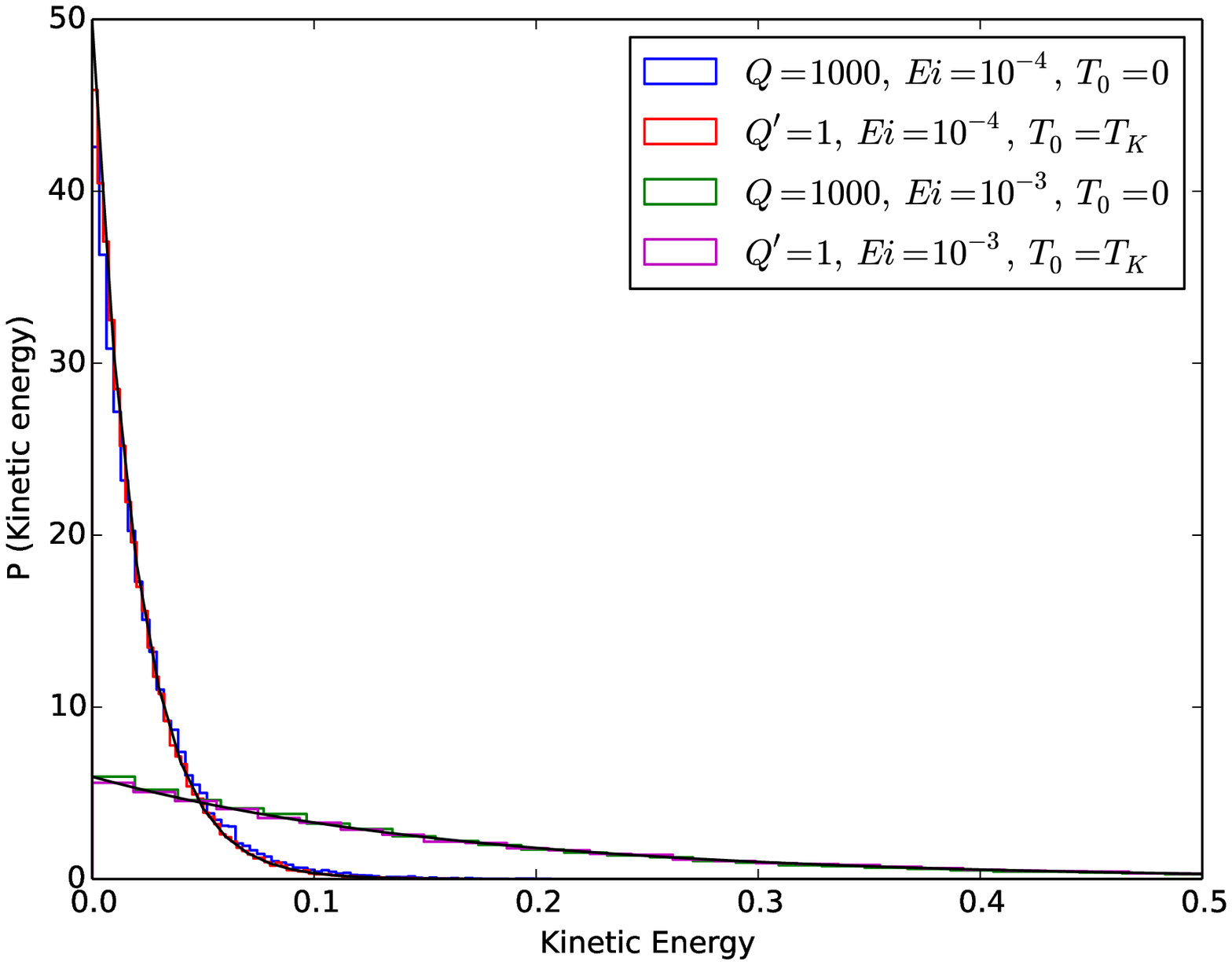}
\par\end{centering}

\caption{\label{fig:ke}Distribution of the kinetic energies $e_K=
\frac{1}{2} m v_i^2$ of the
individual particles in the athermal and thermostatted systems (as indicated
in the legend). Solid black lines are fits to Boltzmann distributions $\propto
e^{-e_K/T_K}$}
\end{figure}

Figure~\ref{fig:ke} presents the distribution of the kinetic energies among the
particles in the system.

\section{III. Robustness of our findings: test of a different thermostat}

\begin{figure}[H]
\begin{centering}
\includegraphics[width=8cm]{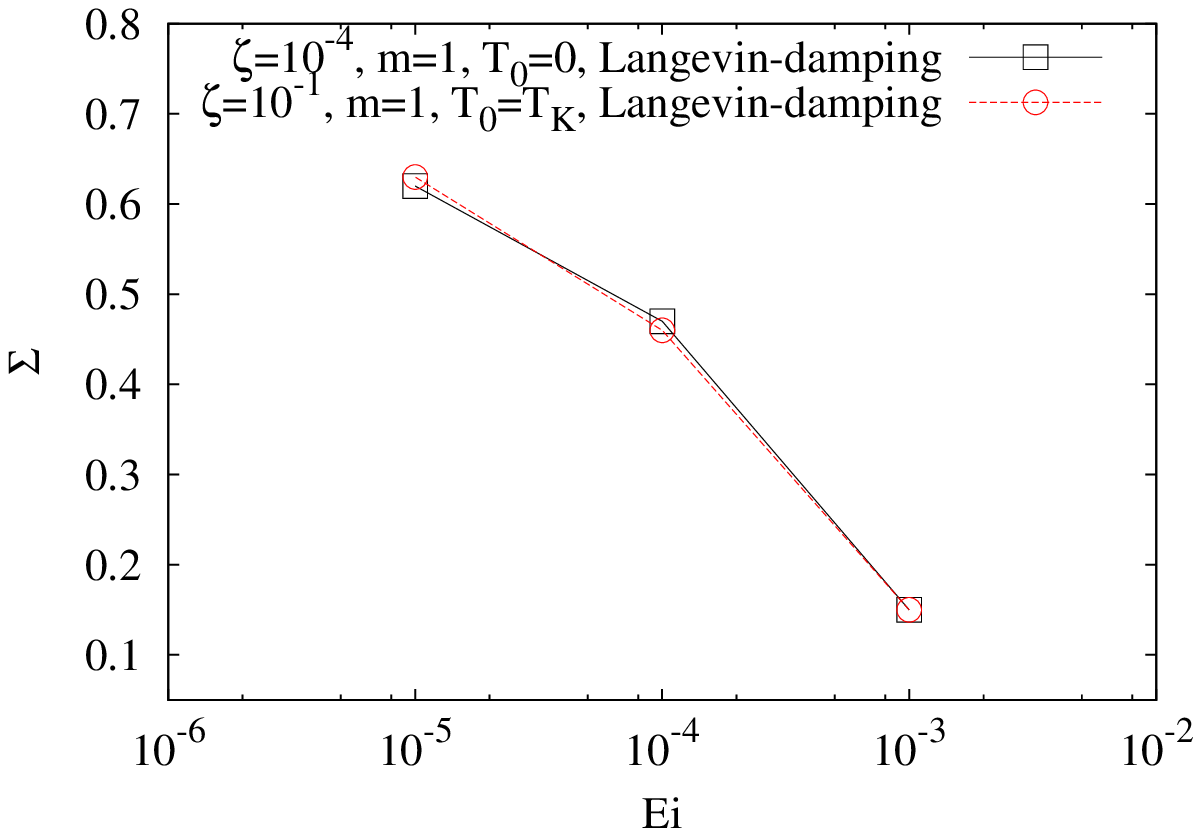}
\par\end{centering}
\caption{\label{fig:Langevin} Flow curves obtained with a Langevin thermostat
(\emph{black})
in an underdamped athermal system and (\emph{red}) in the associated (less
underdamped)
thermostatted
system.}
\end{figure}

To test the robustness of our results, we subsitute a Langevin thermostat for
the Dissipative Particle Dynamics scheme used in the main text. 
Figure~\ref{fig:Langevin} proves that, for low enough damping strength, the
athermal Langevin flow curve also becomes nonmonotonic, with a rate-weakening
region. More importantly, the
thermostatting procedure described in the main text is also operational here,
as evidenced in the figure. This confirms the connection between the
underdamped rheology and the extent of kinetic heating.

\section{IV. Qualitative explanation for the nonmonotonic flow curves in the
highly underdamped regime}

As can be seen in Fig.~2 of the paper, the flow curves at large enough
$\mathrm{Q}$ are nonmonotonic and thus
display a minimum $\left(\mathrm{Ei}_{m},\Sigma_{m}\right)$, where $\mathrm{Ei}$
is the rescaled shear rate. At this point, using the chain rule,
\begin{eqnarray*}
0 & = & \frac{d\Sigma}{d\mathrm{Ei}}\left(\mathrm{Ei}_{m}\right)\\
 & = & \frac{\partial\Sigma}{\partial
T_{K}}\Big|_{\mathrm{Ei}_{m}}\cdot\frac{dT_{K}}{d\mathrm{Ei}}+\frac{
\partial\Sigma}{\partial\mathrm{Ei}}\Big|_{T_{K}=T_{K}\left(\mathrm{Ei}_{m}
\right)},
\end{eqnarray*}
(As a technical detail, note that, in practice, the second partial
derivative - at fixed kinetic temperature - can only be calculated
in a system with stronger damping thermostatted to
$T_{0}=T_{K}\left(\mathrm{Ei}_{m}\right)$).
Thus, the presence of a minimum results from a competition between
two (antagonistic) effects:

(i) an increase of the sample temperature with the shear-rate, which
bends the flow curve downwards

(ii) an intrinsic rate effect at large shear rates, whereby the driving
gets so strong that particles no longer remain at the bottom of the
energy wells (collisions, etc.).

When $\mathrm{Q}$ increases, \emph{i.e.}, when the damping weakens,
the temperature effect (i) gets stronger, so that the minimum is expected
to be deeper, which is indeed confirmed by Fig. 2 of the main text.

\section{V. Deviations from the predicted behaviors observed at high kinetic
temperatures}

At high kinetic temperatures $T_K$(\emph{i.e.}, in particular at large shear
rates and/or large values of Q), Figs.~2 and 3 of the main text show that
deviations appear from the Chattoraj-Lemaitre formula for the
temperature-dependence of the stress (Eq.~3), and, to a much lower extent, from
the scaling prediction represented in Fig.~3.

This is not surprising: Chattoraj and Lemaitre explicitly derived their for
formula in the low-temperature limit, and our scaling prediction was based on
the hypothesis that regions yield (undergo a plastic event) at a local stress
$\Sigma_0$, whereas, owing to thermal activation, the local stress at yield may
be significantly lower at high $T_K$.